\documentclass[reprint,groupedaddress,showpacs,amsmath,amssymb,aps,pra,floatfix]{revtex4-1}
\usepackage{graphicx}
\usepackage{xcolor,colortbl}
\usepackage{array}
\usepackage{bm}
\usepackage{multirow}
\usepackage{makecell}
\usepackage{booktabs}

\begin{document}
\title{Telecom-band Multi-Type Spontaneous Parametric Downconversion in Periodically Polarized Nonlinear Materials}

\author{Xi-Yu Liu${}^{1}$}

\author{Ya-Fei Yu${}^{1}$}

\author{Zheng-Jun Wei${}^{2}$}

\author{Tian-Ming Zhao${}^{1}$}
\email{zhaotm@scnu.edu.cn}  

\author{Jin-Dong Wang${}^{2}$}
\email{wangjindong@m.scnu.edu.cn}  

\affiliation{${}^{1}$Guangdong Provincial Key Laboratory of Nanophotonic Functional Materials and Devices, South China Normal University, Guangzhou, 510631, China\\
${}^{2}$Guangdong Provincial Key Laboratory of Quantum Engineering and Quantum Materials, South China Normal University, Guangzhou, 510631, China\\
}

\date{\today}

\begin{abstract}
Spontaneous parametric downconversion is an essential technique in quantum optics experiments. In this paper, various quasi-phase-matching processes in several typical periodically polarized nonlinear materials are analyzed and calculated. Furthermore, a general method for realizing multiple types of quasi-phase-matching in a monolithic material is presented. Finally, a novel design to prepare multiple entangled photon pairs based on the Sagnac interferometer is discussed. This technology can be applied to tiny optical paths in the telecom \textit{C} band, saving both cost and space.
\end{abstract}

\maketitle


An inevitably pursued research direction in quantum information aims to develop large-scale quantum computing and long-distance quantum communication\cite{GENOVESE2005319}. One of the ongoing efforts is to prepare high-quality quantum light sources in the telecom band, which is required by both long-distance quantum key distribution and multi-node quantum networks\cite{cite-key,PhysRevA.59.4249}. Until now, diverse systems have been implemented to prepare quantum light sources ranging from nonlinear materials and atomic ensembles to semiconductor quantum dots(QDs) and nitrogen-vacancy(NV) color centers\cite{Hanson}. Under communication wavelengths, however, quasi-phase-matching (QPM) spontaneous parametric downconversion  (SPDC) in the nonlinear optical system is the most mature and widely used technique because of its sources' high brightness and visibility\cite{PhysRev.127.1918}. Commonly used nonlinear materials for QPM are PPKTP, PPLN, and PPSLT.

\begin{figure}[h]
	\includegraphics[width=1\columnwidth]{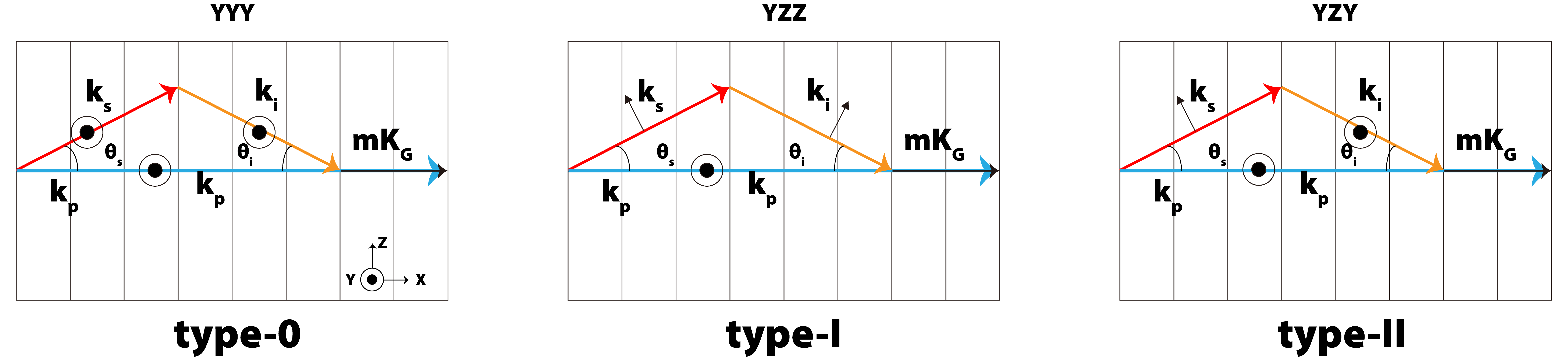}
	\caption{(color online) Three types of QPM processes in periodically poled nonlinear materials. }
	\label{types}
\end{figure} 

SPDC is a second-order nonlinear effect in optical materials. When a high-frequency pump photon enters the material, it splits into a low-frequency signal photon and idle photon pairs at quasi-phase-matching(QPM)\cite{Schaeff:12,Lee:15,Steinlechner:14}. According to the different polarizations of the pump photon and the two down-converted photons, there are three types of SPDC called type-0, type-I, and type-II. As shown in Fig. \ref{types}, in type-0 SPDC, the signal photon, idler photon, and pump photon have the same polarization. In type-I SPDC, the signal photon has the same polarization as the idler photon, but both are perpendicular to the pump. Finally, in type-II SPDC, the polarization of the signal photon and idle photon is perpendicular.

Various methods have been developed to improve the quality of entangled photon sources. One is an in-depth discussion of phase-matching. It is hoped that through the comparative study of different phase-matching types, conditions more suitable for the preparation of quantum light sources will be discovered, such as the realization of multiple types of SPDC in a monolithic material. In 2009, NIST designed a PPKTP waveguide that simultaneously realizes type-0 and type-II SPDC and compared their characteristics\cite{Chen:09}. In 2012, PNU designed a PPKTP crystal to realize both type-0 and type-II SPDC processes at 800 nm\cite{Lee:2012aa}.These works provide a theoretical and experimental basis for the simultaneous realization of multiple types of SPDC in the same nonlinear material. However, the current working wavelength is limited to around 800 nm. Entanglement distribution can be conveniently established at telecom wavelengths, where optical fibers are less lossy and conventional polarization entanglement sources are principally closer to ideal Bell states, in the context of city-scale optical fiber-based networks\cite{Jin}. In 2021, UA demonstrated that phase matching for both type-I and type-II can be realized simultaneously in the same waveguide at 1550 nm, but correlated photons generated from type-II SPDC was overlapped with type-I SPDC in frequency\cite{Briggs:21}. 

In this work, we calculate and analyze the characteristics of different types of SPDC in the telecom \textit{C} band. Furthermore, we investigate the parameters and their potential applications for the simultaneous realization of multi-type of SPDC in several typical nonlinear materials. Finally, we designed the new experiment to generate multi-pair and multi-photon entangled light sources.


We first analyze various types of SPDC (type-0, type-I, and type-II) in three typical periodically poled nonlinear materials (PPKTP, PPLN, and PPLT). The process satisfies the following QPM conditions:
\begin{eqnarray}\label{eq1}
\omega_{p}=\omega_{s}+\omega_{i} (\text{or} \frac{1}{\lambda_{p}}=\frac{1}{\lambda_{s}}+\frac{1}{\lambda_{i}})
\end{eqnarray}
\begin{eqnarray}\label{eq2}
\vec{k}_{p}=\vec{k}_{s}+\vec{k}_{i}+m\vec{K}_{g} (K_{g}=\frac{2\pi}{\Lambda})
\end{eqnarray}
\begin{eqnarray}\label{eq3}
\vert\vec{k}_{j}\vert=\vert\frac{2\pi\vec{n}_{j}}{\lambda_{j}}\vert(j=p, s, i)
\end{eqnarray}
where the superscripts p, s and i represent the pump, the signal and the idler, respectively. In Eq.(\ref{eq1}), $\omega$ is the angular frequency, and $\lambda$ is the wavelength. Here we select the pump wavelength as 775 nm, and that of the signal and idler are both 1550 nm. Eq.(\ref{eq2}) means the conservation of momentum, where $\vec{k}$ is the wave vector, $K_{g}$ is the periodic polarization vector, $\Lambda$ is the poling period of the material, and m, an integer, is referred to as QPM order. In Eq.(\ref{eq3}), n is the refractive index of the light, which varies with wavelength, temperature, and angle. And considering the angle, the above QPM equations can be expressed as,

\begin{eqnarray}\label{eq4}
k_{s}\cos{\theta_{s}}+k_{i}\cos{\theta_{i}}=k_{p}-mK_{g}
\end{eqnarray}
\begin{eqnarray}\label{eq5}
k_{s}\sin{\theta_{s}}=k_{i}\sin{\theta_{i}}
\end{eqnarray}

We can see the refractive index's effect on phase matching from the above formulas. According to the Sellmeier equation, the refractive index corresponds to different polarization of photons in nonlinear materials, which varies by type of SPDC\cite{Gayer:2008aa}. On the other hand, as we know, the two photons generated by type-II SPDC can be separated by polarization. In contrast, the polarization of photons generated under type-0 and type-I are indistinguishable. However, they can be separated from the space for non-collinear QPM. So in this paper, our calculations are all for non-collinear type-0, type-I, and collinear type-II SPDC. In this case, the phase matching and refractive index are calculated by the following equations.

(1)Non-collinear Type-0 SPDC
\begin{eqnarray}\label{eq6}
n_{s(i)}(T, \lambda_{s(i)})=n_{y}(T, \lambda_{p})
\end{eqnarray}
\begin{eqnarray}\label{eq7}
n_{p}=n_{y}(T, \lambda_{p})
\end{eqnarray}

(2)Non-collinear Type-I SPDC
\begin{eqnarray}\label{eq8}
\begin{split}
&n_{s(i)}(\theta_{s(i)}, T, \lambda_{s(i)})\\
=&\frac{n_{x}(T, \lambda_{s(i)})n_{z}(T, \lambda_{s(i)})}{\sqrt{n_{x}^{2}(T, \lambda_{s(i)})\cos^{2}{\theta_{s(i)}}+n_{z}^{2}(T, \lambda_{s(i)})\sin^{2}{\theta_{s(i)}}}}\\
\end{split}
\end{eqnarray}
\begin{eqnarray}\label{eq9}
n_{p}=n_{y}(T, \lambda_{p})
\end{eqnarray}

(3)Collinear Type-II SPDC
\begin{eqnarray}\label{eq10}
\begin{split}
&n_{s}(\theta_{s}, T, \lambda_{s})\\
=&\frac{n_{x}(T, \lambda_{s})n_{z}(T, \lambda_{s})}{\sqrt{n_{x}^{2}(T, \lambda_{s})\cos^{2}{\theta_{s}}+n_{z}^{2}(T, \lambda_{s})\sin^{2}{\theta_{s}}}}\\
\end{split}
\end{eqnarray}
\begin{eqnarray}\label{eq11}
n_{p}=n_{y}(T, \lambda_{p})
\end{eqnarray}
\begin{eqnarray}\label{eq12}
n_{i}=n_{y}(T, \lambda_{i})
\end{eqnarray}

The beams propagate in the x-z plane and $\theta_{s(i)}$ is the angle between $\vec{k}_{s(i)}$ and $\vec{k}_{p}$ in the x-y plane(see Fig.\ref{types}). \textit{T} is the crystal temperature.
\begin{figure}[h]
	\includegraphics[width=1\columnwidth]{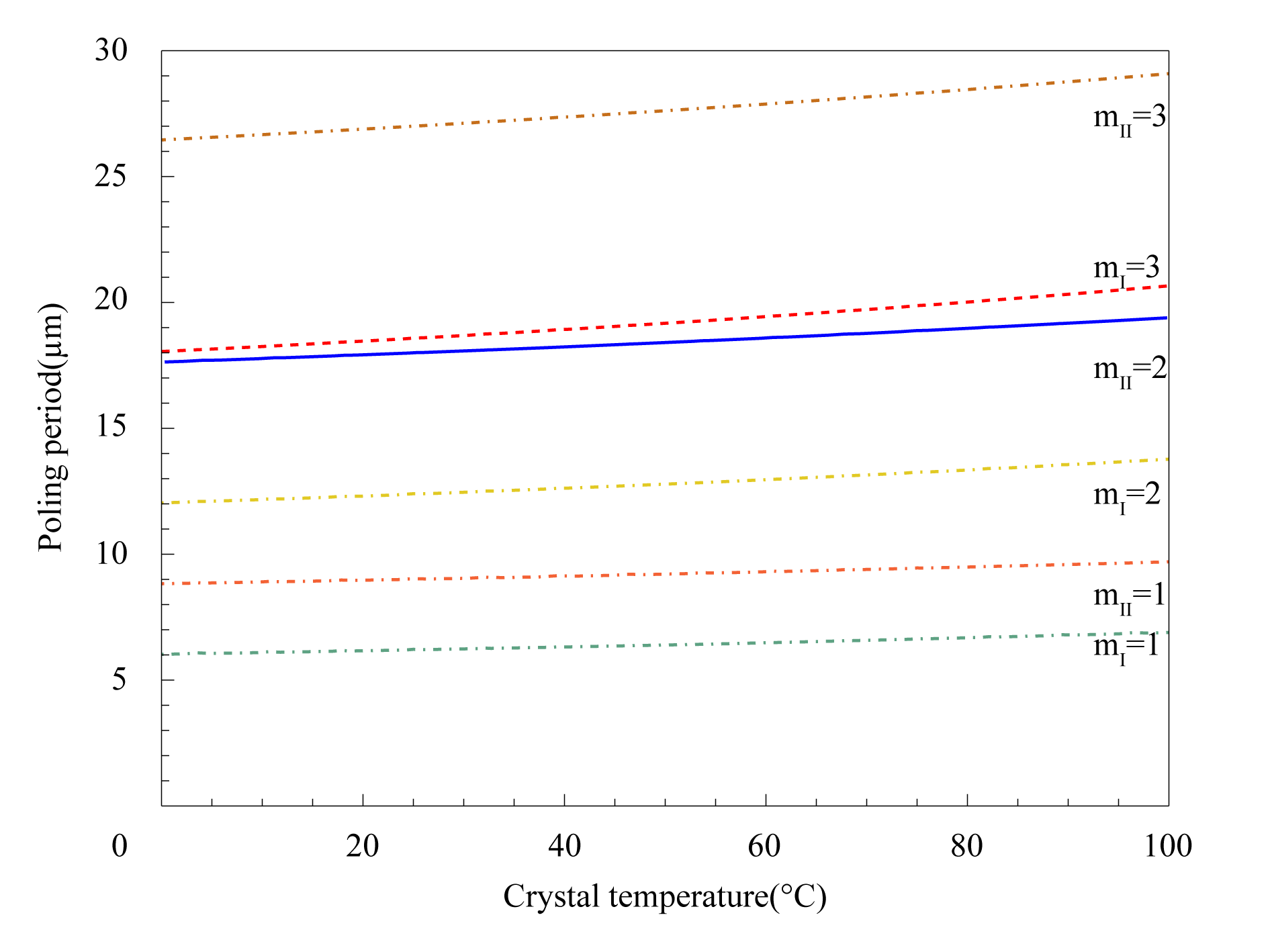}
	\caption{(color online) Dependence of poling period on crystal temperature. $\text{m}_{\text{I}}$ and $\text{m}_{\text{II}}$ represent orders of type-I and type-II SPDC respectively. Solid blue curve: for second-order type-II, and dashed red curve: for third-order type-I.}
	\label{order}
\end{figure} 

Next, we find conditions for the simultaneous realization of multiple types of QPM in the same nonlinear material. Without loss of generality, take PPLN as an example. We first study the effect of the QPM order. In the QPM process, after the horizontally polarized pump with a wavelength of 775 nm is injected into the PPLN crystal, a degenerate photon pair with a wavelength of 1550 nm is obtained from SPDC. We show the curve between the crystal temperature and the poling period of 1-3 order type-I(o$\rightarrow$e+e) and type-II(o$\rightarrow$o+e) SPDC under collinear conditions in Fig.\ref{order}.

\begin{figure}[h]
	\includegraphics[width=1\columnwidth]{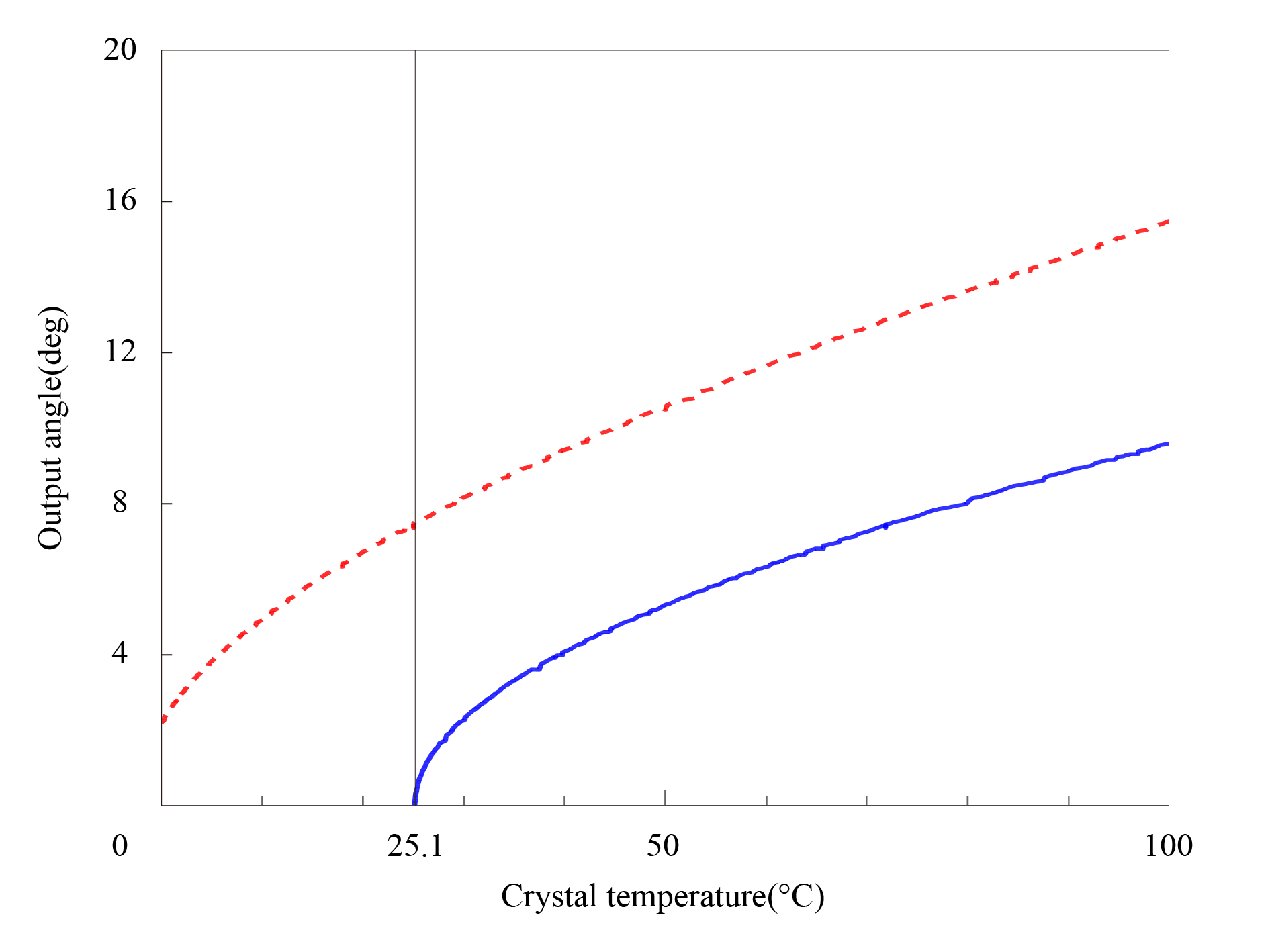}
	\caption{(color online) Dependence of Output angle on crystal temperature. Solid blue curve: for type-II, dashed red curve: for type-I.}
	\label{angle}
\end{figure} 

Fig.\ref{order} shows that collinear third-order type-I (dashed red line) and second-order type-II (solid blue line) curves are closely adjacent without intersecting points. So we can not find out both types of SPDC at the same time directly. Instead, we can change type-I from collinear to non-collinear and select an appropriate angle. Moreover, the type-II curve in Fig.\ref{order}  shows that the crystal temperature is 17-19 $\mu$m. So we may as well take the period as 18.000 $\mu$m. Then we get the relationships between the crystal temperature and the output angle of the signal satisfying QPM, as shown in Fig.\ref{angle}. Again, the solid blue line is for type-II, and the dashed red line is for type-I. At a temperature of 25.1$^{\circ}$C, the exit angle of the blue line calculated by the Snell's law is zero, meaning collinear. Meanwhile, under 25.1$^{\circ}$C, non-collinear type-I QPM can be achieved with an exit angle of 7$^{\circ}$.

In conclusion, it can perform second-order type-II SPDC collinearly and third-order type-I SPDC with an exit angle of 7$^{\circ}$ at the same time when the PPLN crystal has a poling period of 18.000 $\mu$m and a temperature of 25.1$^{\circ}$C.

\begin{figure}[h]
	\includegraphics[width=1\columnwidth]{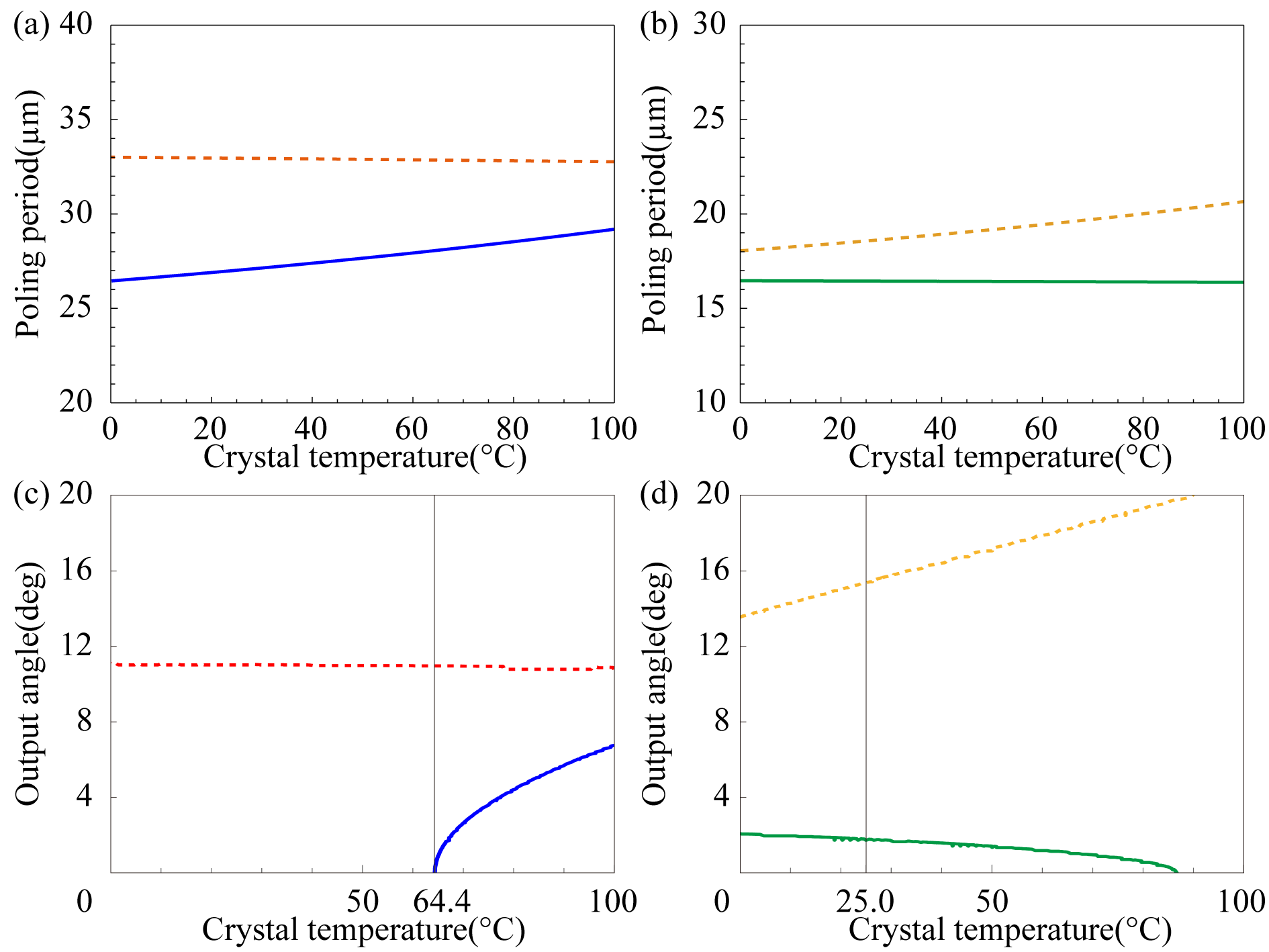}
	\caption{(color online) (a) and (b) Dependence of poling period on crystal temperature. (c) and (d) Dependence of Output angle on crystal temperature. Solid blue curve: for type-II QPM, and dashed yellow curve: for type-I QPM. Both of solid green curve and dashed red curve are for type-0 QPM.}
	\label{4pic}
\end{figure}
\begin{figure}[h]
	\includegraphics[width=1\columnwidth]{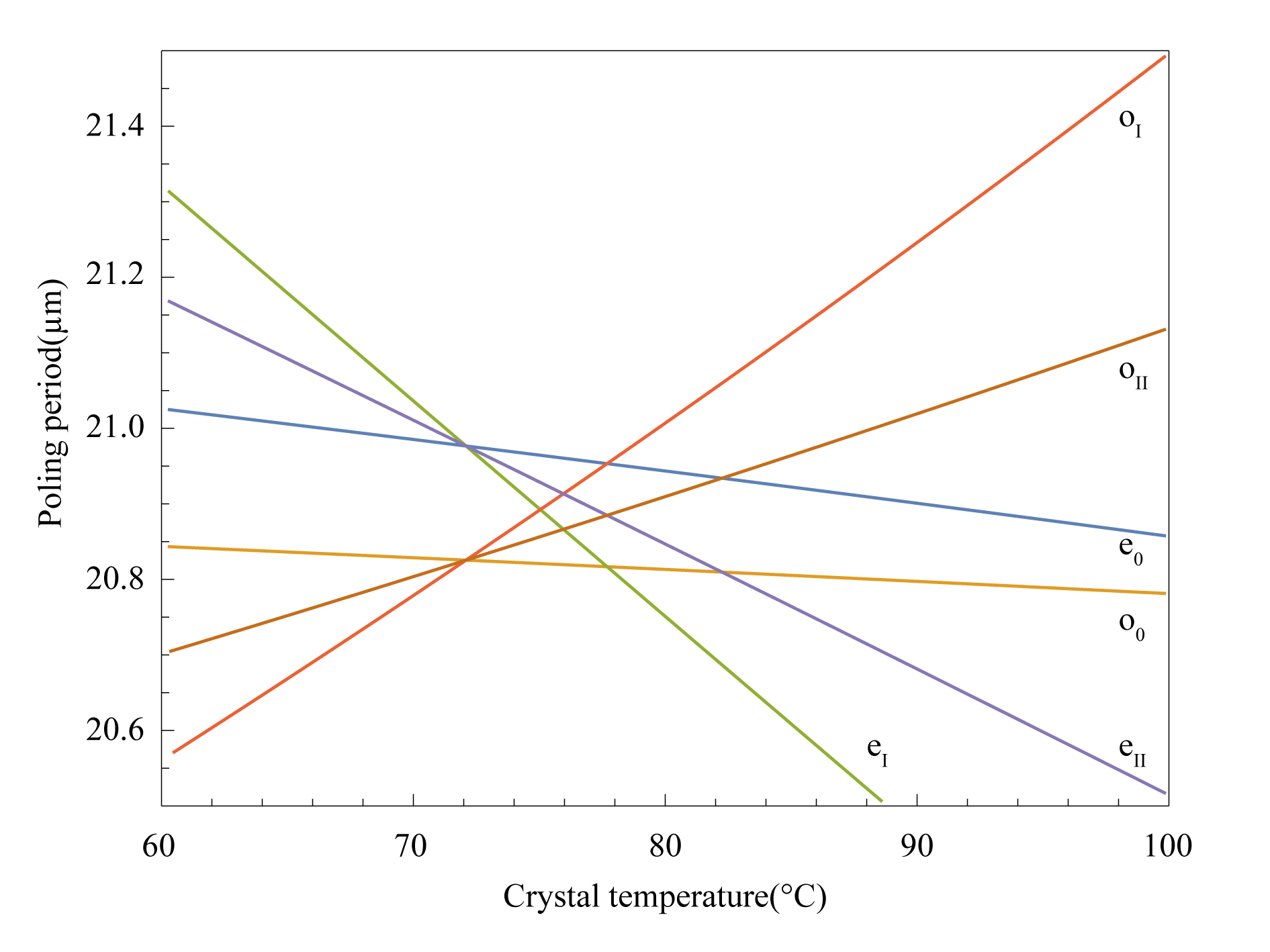}
	\caption{(color online) Dependence of poling period on the crystal temperature. The letters o and e represent the ordinary and  extraordinary components of light, and subscripts 0, I and II indicate type-0, type-I and type-II, respectively.}
	\label{ppslt}
\end{figure}

 
The simulation results for the other two situations are shown in Fig.\ref{4pic}, where Fig.\ref{4pic} (a) and (c) are for type-0(o$\rightarrow$o+o) and type-II(o$\rightarrow$o+e) QPM. When the PPLN crystal has a poling period of 28.0 $\mu$m and a temperature of 64.4$^{\circ}$C, the non-collinear second-order type-0 and collinear third-order type-II SPDC processes can be simultaneously performed. Similarly, Fig.\ref{4pic} (b) and (d) are for type-0(o$\rightarrow$o+o) and type-I(o$\rightarrow$e +e) QPM. The double-QPM condition can be satisfied with a poling period of 16.4 $\mu$m and a temperature of 25$^{\circ}$C. We summarize all the results in Table \ref{table1}.
\begin{table}[t]
	\centering 
     \renewcommand{\arraystretch}{2.5}
	\caption{Parameters}  
	\label{table1} 
	\scalebox{0.8}{
	\begin{tabular}{m{1.4cm}<{\centering} m{1.5cm}<{\centering} m{1.3cm}<{\centering} m{2.3cm}<{\centering} m{1cm}<{\centering} m{1.4cm}<{\centering}} 
	        \toprule[2pt]  		
		\textbf{Material}&\textbf{Temp.}&\makecell{\textbf{Poling}\\\textbf{period}} & \textbf{Process} & \textbf{Order}&\makecell{\textbf{Output}\\\textbf{angle}}\\
           \\[-22pt]\midrule[1pt]
		\multirow{3}{*}{PPLN} & 64.4$^{\circ}$C & 28.002$\mu$m & \makecell{type-II:o$\rightarrow$e+o\\type-0:o$\rightarrow$o+o}& \makecell{3\\2}&\makecell{0$^{\circ}$\\11$^{\circ}$}\\[2pt] 
		 & 25.1$^{\circ}$C & 18.000$\mu$m & \makecell{type-II:o$\rightarrow$e+o\\type-I:o$\rightarrow$e+e}& \makecell{2\\3}&\makecell{0$^{\circ}$\\7$^{\circ}$} \\[2pt]  
		 & 25.0$^{\circ}$C & 16.400$\mu$m & \makecell{type-0:o$\rightarrow$o+o\\type-I:o$\rightarrow$e+e}& \makecell{1\\3}    &\makecell{2$^{\circ}$\\15$^{\circ}$}  \\[16pt] 
		 \multirow{2}{*}[-10pt]{PPSLT} & \multirow{2}{*}[-10pt]{72.1$^{\circ}$C} & 20.826$\mu$m & \makecell{type-0:o$\rightarrow$o+o\\type-I:o$\rightarrow$e+e\\type-II:o$\rightarrow$o+e}& \makecell{1\\1\\1}&\makecell{0$^{\circ}$\\0$^{\circ}$\\0$^{\circ}$}\\[16pt] 
		 &  & 20.978$\mu$m & \makecell{type-0:e$\rightarrow$e+e\\type-I:e$\rightarrow$o+o\\type-II:e$\rightarrow$e+o}& \makecell{1\\1\\1}&\makecell{0$^{\circ}$\\0$^{\circ}$\\0$^{\circ}$}\\[2pt]

           \\[-22pt]\bottomrule[2pt]
	\end{tabular}}
\end{table}
The above method is also applicable to other nonlinear materials. Furthermore, for the material of PPSLT, we find two conditions to realize triple-type QPM as well\cite{Dolev:2009aa}. As shown in Fig.\ref{ppslt}, we plot the relationship between temperature and poling period under the three types of phase matching when the pump light is o light and e light, respectively. Two intersection points represent two triple-type QPM conditions. Both intersection temperatures are 72.1$^{\circ}$C, and the poling periods are 20.83 $\mu$m and 20.98 $\mu$m, respectively.

\begin{figure}[t]
	\includegraphics[width=1\columnwidth]{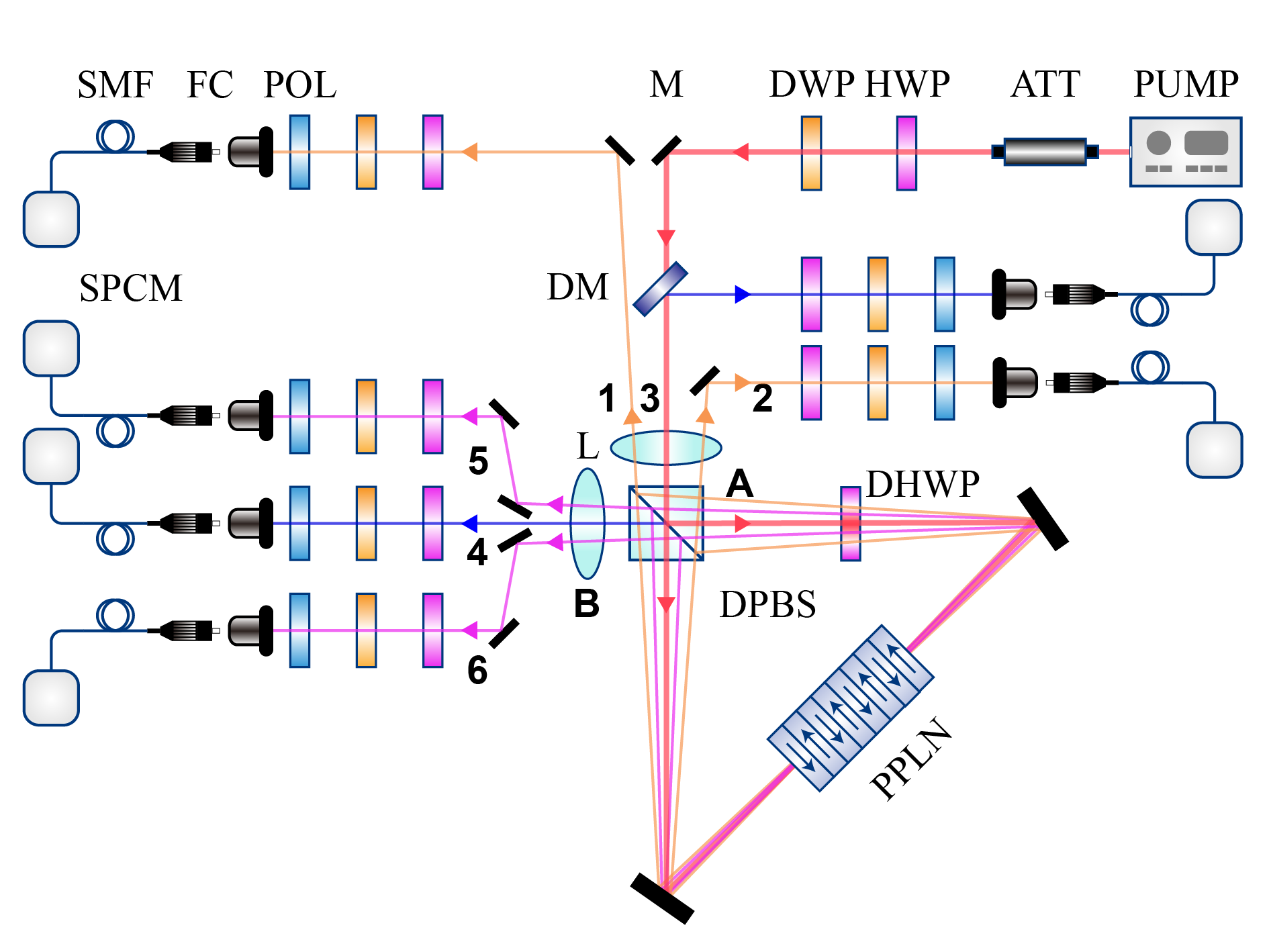}
	\caption{(color online) Schematic setup of the Sagnac interferometer. A 775-nm pump beam, with a 45$^{\circ}$ polarization, is focused to the center of the PPLN crystal placed at the center of the Sagnac loop. The down-converted photons propagating in path 1-6 output from port A or B of PBS are coupled into single-mode fibers (SMFs). The DM is a dichroic mirror. And the DHWP is a dual-wavelength half-wave plate.}
	\label{inter}
\end{figure}

In the previous section, we discuss how to achieve multiple types of QPM in the same nonlinear material. Next, we talk about a possible experimental application of this technique. Fig.\ref{inter} shows the experimental setup for generating polarization-entangled photon sources based on the Sagnac interferometer. When the pump light passes through a PPLN crystal with a poling period of 64.4$\mu$m at a temperature of 28.0$^{\circ}$C, it is known that two types of SPDC processes, the non-collinear type-0 and the collinear type-II, can be simultaneously implemented. Firstly, let's talk about the non-collinear type-0 process. A 45$^{\circ}$ polarized pump beam is split into two beams by PBS. The clockwise vertically polarized light is converted into horizontally polarized light by a fixed 45$^{\circ}$ HWP, and then generates horizontally polarized photon pairs:$|\text{V}\rangle_{p}\stackrel{\text{DHWP}}{\longrightarrow} |\text{H}\rangle_{p}\stackrel{\text{PPLN}}{\longrightarrow} |\text{H}\rangle_{1}|\text{H}\rangle_{2}$. The photon pairs exit from port A of the PBS. Meanwhile, the counterclockwise horizontally polarized light passes through the PPLN crystal to generate horizontally polarized photon pairs, which are then converted into vertically polarized ones through the HWP: $|\text{H}\rangle_{p}\stackrel{\text{PPLN}}{\longrightarrow} |\text{H}\rangle_{1}|\text{H}\rangle_{2}\stackrel{\text{DHWP}}{\longrightarrow} |\text{V}\rangle_{1}|\text{V}\rangle_{2}$. The photons also exit from the port A. Therefore, we get the entangled satate:$|\Phi\rangle=\frac{1}{\sqrt{2}}(|\text{H}\rangle_{1}|\text{H}\rangle_{2}+e^{i\phi_{0}}|\text{V}\rangle_{1}|\text{V}\rangle_{2})$, where the subscripts represent different spatial paths, and $\phi_{0}$ represents the phase between the two-photon pairs. On this basis, the Bell state $|\Phi^{+}\rangle=\frac{1}{\sqrt{2}}(|\text{H}\rangle_{1}|\text{H}\rangle_{2}+|\text{V}\rangle_{1}|\text{V}\rangle_{2})$ can be obtained by adjusting the phase compensation in the optical path after the interferometer so that $\phi_{0}=0$. In addition, a new Bell state can also be obtained by rotating the HWP by 45$^{\circ}$: $|\Psi\rangle=\frac{1}{\sqrt{2}}(|\text{H}\rangle_{1}|\text{V}\rangle_{2}+e^{i\phi_{0}}|\text{V}\rangle_{1}|\text{H}\rangle_{2})$. The 1-QWP can be used to adjust $\phi_{0}=\pi$ by rotating the angle by 90$^{\circ}$, so as to obtain $\Phi^{-}$ and $\Psi^{-}$ states. Secondly, we discuss the collinear type-II process. Two-photon pairs generated by the type-II SPDC in the PPLN crystal. The polarization-entangled state is $|\Phi\rangle=\frac{1}{\sqrt{2}}(|\text{H}\rangle_{3}|\text{V}\rangle_{4}+e^{i\phi_{II}}|\text{V}\rangle_{3}|\text{H}\rangle_{4})$, which can be changed to different Bell states by adjusting the HWP and QWP. Finally, in the non-collinear type-I process, the two-photon pairs both exit from the B port of the PBS and are successfully prepared as polarization-entangled state $|\Phi\rangle=\frac{1}{\sqrt{2}}(|\text{V}\rangle_{5}|\text{V}\rangle_{6}+e^{i\phi_{I}}|\text{H}\rangle_{5}|\text{H}\rangle_{6})$. Different Bell states can also be obtained by adjusting the HWP or QWP.

There are still a few technical difficulties to overcome. For example, it is complicacy to maintain a uniform poling period. As type-II phase matching in Table \ref{table1} does depend on period poling, a tiny change in the poling period ($\Lambda$ = 18.000$\mu$m, $\nabla \Lambda$ = 4nm, $\frac{\nabla \Lambda}{\Lambda} \approx$ 0.0002) can alter the wave vector mismatch. For the purpose of compensating the mismatch, we can set the crystal temperature to 25.4$^{\circ}$C. However, based on this experiment setup, two or three entanglement states can be generated simultaneously using only one nonlinear material. Compared with the experiments that require a set of optical systems to prepare multiple entangled photon pairs, it saves space and cost\cite{PhysRevLett.128.040402, Ho:2022aa}.


This paper gives the theoretical models and simulation results of multi-type SPDC in periodically polarized nonlinear materials. Furthermore, the optimal conditions for the quantum entanglement source in the telecom \textit{C} band are studied. As a result, the brightness and quality of quantum entangled light sources will be improved through experimental comparison. Besides, we discovered a new method for simultaneously preparing multiple pairs of quantum entangled photons by multi-type QPM. This work will promote quantum information to be practical and widely used in various fields of quantum information, such as quantum key distribution, quantum teleportation, quantum computing, and quantum simulation. Our results pave the way for entanglement distribution over organized fibers at metropolitan scale.

 \section*{Acknowledgements}
Zhao is supported by NSFC grant (No.11904422). 


\bibliographystyle{apsrev4-1}
\bibliography{types_refs}

\begin{thebibliography}{16}%
\makeatletter
\providecommand \@ifxundefined [1]{%
 \@ifx{#1\undefined}
}%
\providecommand \@ifnum [1]{%
 \ifnum #1\expandafter \@firstoftwo
 \else \expandafter \@secondoftwo
 \fi
}%
\providecommand \@ifx [1]{%
 \ifx #1\expandafter \@firstoftwo
 \else \expandafter \@secondoftwo
 \fi
}%
\providecommand \natexlab [1]{#1}%
\providecommand \enquote  [1]{``#1''}%
\providecommand \bibnamefont  [1]{#1}%
\providecommand \bibfnamefont [1]{#1}%
\providecommand \citenamefont [1]{#1}%
\providecommand \href@noop [0]{\@secondoftwo}%
\providecommand \href [0]{\begingroup \@sanitize@url \@href}%
\providecommand \@href[1]{\@@startlink{#1}\@@href}%
\providecommand \@@href[1]{\endgroup#1\@@endlink}%
\providecommand \@sanitize@url [0]{\catcode `\\12\catcode `\$12\catcode
  `\&12\catcode `\#12\catcode `\^12\catcode `\_12\catcode `\%12\relax}%
\providecommand \@@startlink[1]{}%
\providecommand \@@endlink[0]{}%
\providecommand \url  [0]{\begingroup\@sanitize@url \@url }%
\providecommand \@url [1]{\endgroup\@href {#1}{\urlprefix }}%
\providecommand \urlprefix  [0]{URL }%
\providecommand \Eprint [0]{\href }%
\providecommand \doibase [0]{http://dx.doi.org/}%
\providecommand \selectlanguage [0]{\@gobble}%
\providecommand \bibinfo  [0]{\@secondoftwo}%
\providecommand \bibfield  [0]{\@secondoftwo}%
\providecommand \translation [1]{[#1]}%
\providecommand \BibitemOpen [0]{}%
\providecommand \bibitemStop [0]{}%
\providecommand \bibitemNoStop [0]{.\EOS\space}%
\providecommand \EOS [0]{\spacefactor3000\relax}%
\providecommand \BibitemShut  [1]{\csname bibitem#1\endcsname}%
\let\auto@bib@innerbib\@empty
\bibitem [{\citenamefont {Genovese}(2005)}]{GENOVESE2005319}%
  \BibitemOpen
  \bibfield  {author} {\bibinfo {author} {\bibfnamefont {M.}~\bibnamefont
  {Genovese}},\ }\href {\doibase https://doi.org/10.1016/j.physrep.2005.03.003}
  {\bibfield  {journal} {\bibinfo  {journal} {Physics Reports}\ }\textbf
  {\bibinfo {volume} {413}},\ \bibinfo {pages} {319} (\bibinfo {year}
  {2005})}\BibitemShut {NoStop}%
\bibitem [{\citenamefont {O'Brien}\ \emph {et~al.}(2009)\citenamefont
  {O'Brien}, \citenamefont {Furusawa},\ and\ \citenamefont {Vu{\v
  c}kovi{\'c}}}]{cite-key}%
  \BibitemOpen
  \bibfield  {author} {\bibinfo {author} {\bibfnamefont {J.~L.}\ \bibnamefont
  {O'Brien}}, \bibinfo {author} {\bibfnamefont {A.}~\bibnamefont {Furusawa}}, \
  and\ \bibinfo {author} {\bibfnamefont {J.}~\bibnamefont {Vu{\v
  c}kovi{\'c}}},\ }\href {\doibase 10.1038/nphoton.2009.229} {\bibfield
  {journal} {\bibinfo  {journal} {Nature Photonics}\ }\textbf {\bibinfo
  {volume} {3}},\ \bibinfo {pages} {687} (\bibinfo {year} {2009})}\BibitemShut
  {NoStop}%
\bibitem [{\citenamefont {Cirac}\ \emph {et~al.}(1999)\citenamefont {Cirac},
  \citenamefont {Ekert}, \citenamefont {Huelga},\ and\ \citenamefont
  {Macchiavello}}]{PhysRevA.59.4249}%
  \BibitemOpen
  \bibfield  {author} {\bibinfo {author} {\bibfnamefont {J.~I.}\ \bibnamefont
  {Cirac}}, \bibinfo {author} {\bibfnamefont {A.~K.}\ \bibnamefont {Ekert}},
  \bibinfo {author} {\bibfnamefont {S.~F.}\ \bibnamefont {Huelga}}, \ and\
  \bibinfo {author} {\bibfnamefont {C.}~\bibnamefont {Macchiavello}},\ }\href
  {\doibase 10.1103/PhysRevA.59.4249} {\bibfield  {journal} {\bibinfo
  {journal} {Phys. Rev. A}\ }\textbf {\bibinfo {volume} {59}},\ \bibinfo
  {pages} {4249} (\bibinfo {year} {1999})}\BibitemShut {NoStop}%
\bibitem [{\citenamefont {Bernien}\ \emph {et~al.}(2013)\citenamefont
  {Bernien}, \citenamefont {Hensen}, \citenamefont {Pfaff}, \citenamefont
  {Koolstra}, \citenamefont {Blok}, \citenamefont {Robledo}, \citenamefont
  {Taminiau}, \citenamefont {Markham}, \citenamefont {Twitchen}, \citenamefont
  {Childress},\ and\ \citenamefont {Hanson}}]{Hanson}%
  \BibitemOpen
  \bibfield  {author} {\bibinfo {author} {\bibfnamefont {H.}~\bibnamefont
  {Bernien}}, \bibinfo {author} {\bibfnamefont {B.}~\bibnamefont {Hensen}},
  \bibinfo {author} {\bibfnamefont {W.}~\bibnamefont {Pfaff}}, \bibinfo
  {author} {\bibfnamefont {G.}~\bibnamefont {Koolstra}}, \bibinfo {author}
  {\bibfnamefont {M.~S.}\ \bibnamefont {Blok}}, \bibinfo {author}
  {\bibfnamefont {L.}~\bibnamefont {Robledo}}, \bibinfo {author} {\bibfnamefont
  {T.~H.}\ \bibnamefont {Taminiau}}, \bibinfo {author} {\bibfnamefont
  {M.}~\bibnamefont {Markham}}, \bibinfo {author} {\bibfnamefont {D.~J.}\
  \bibnamefont {Twitchen}}, \bibinfo {author} {\bibfnamefont {L.}~\bibnamefont
  {Childress}}, \ and\ \bibinfo {author} {\bibfnamefont {R.}~\bibnamefont
  {Hanson}},\ }\href {\doibase 10.1038/nature12016} {\bibfield  {journal}
  {\bibinfo  {journal} {Nature}\ }\textbf {\bibinfo {volume} {497}},\ \bibinfo
  {pages} {86} (\bibinfo {year} {2013})}\BibitemShut {NoStop}%
\bibitem [{\citenamefont {Armstrong}\ \emph {et~al.}(1962)\citenamefont
  {Armstrong}, \citenamefont {Bloembergen}, \citenamefont {Ducuing},\ and\
  \citenamefont {Pershan}}]{PhysRev.127.1918}%
  \BibitemOpen
  \bibfield  {author} {\bibinfo {author} {\bibfnamefont {J.~A.}\ \bibnamefont
  {Armstrong}}, \bibinfo {author} {\bibfnamefont {N.}~\bibnamefont
  {Bloembergen}}, \bibinfo {author} {\bibfnamefont {J.}~\bibnamefont
  {Ducuing}}, \ and\ \bibinfo {author} {\bibfnamefont {P.~S.}\ \bibnamefont
  {Pershan}},\ }\href {\doibase 10.1103/PhysRev.127.1918} {\bibfield  {journal}
  {\bibinfo  {journal} {Phys. Rev.}\ }\textbf {\bibinfo {volume} {127}},\
  \bibinfo {pages} {1918} (\bibinfo {year} {1962})}\BibitemShut {NoStop}%
\bibitem [{\citenamefont {Schaeff}\ \emph {et~al.}(2012)\citenamefont
  {Schaeff}, \citenamefont {Polster}, \citenamefont {Lapkiewicz}, \citenamefont
  {Fickler}, \citenamefont {Ramelow},\ and\ \citenamefont
  {Zeilinger}}]{Schaeff:12}%
  \BibitemOpen
  \bibfield  {author} {\bibinfo {author} {\bibfnamefont {C.}~\bibnamefont
  {Schaeff}}, \bibinfo {author} {\bibfnamefont {R.}~\bibnamefont {Polster}},
  \bibinfo {author} {\bibfnamefont {R.}~\bibnamefont {Lapkiewicz}}, \bibinfo
  {author} {\bibfnamefont {R.}~\bibnamefont {Fickler}}, \bibinfo {author}
  {\bibfnamefont {S.}~\bibnamefont {Ramelow}}, \ and\ \bibinfo {author}
  {\bibfnamefont {A.}~\bibnamefont {Zeilinger}},\ }\href {\doibase
  10.1364/OE.20.016145} {\bibfield  {journal} {\bibinfo  {journal} {Opt.
  Express}\ }\textbf {\bibinfo {volume} {20}},\ \bibinfo {pages} {16145}
  (\bibinfo {year} {2012})}\BibitemShut {NoStop}%
\bibitem [{\citenamefont {Lee}\ \emph {et~al.}(2015)\citenamefont {Lee},
  \citenamefont {Kim}, \citenamefont {Cha},\ and\ \citenamefont
  {Moon}}]{Lee:15}%
  \BibitemOpen
  \bibfield  {author} {\bibinfo {author} {\bibfnamefont {H.~J.}\ \bibnamefont
  {Lee}}, \bibinfo {author} {\bibfnamefont {H.}~\bibnamefont {Kim}}, \bibinfo
  {author} {\bibfnamefont {M.}~\bibnamefont {Cha}}, \ and\ \bibinfo {author}
  {\bibfnamefont {H.~S.}\ \bibnamefont {Moon}},\ }\href {\doibase
  10.1364/OE.23.014203} {\bibfield  {journal} {\bibinfo  {journal} {Opt.
  Express}\ }\textbf {\bibinfo {volume} {23}},\ \bibinfo {pages} {14203}
  (\bibinfo {year} {2015})}\BibitemShut {NoStop}%
\bibitem [{\citenamefont {Steinlechner}\ \emph {et~al.}(2014)\citenamefont
  {Steinlechner}, \citenamefont {Gilaberte}, \citenamefont {Jofre},
  \citenamefont {Scheidl}, \citenamefont {Torres}, \citenamefont {Pruneri},\
  and\ \citenamefont {Ursin}}]{Steinlechner:14}%
  \BibitemOpen
  \bibfield  {author} {\bibinfo {author} {\bibfnamefont {F.}~\bibnamefont
  {Steinlechner}}, \bibinfo {author} {\bibfnamefont {M.}~\bibnamefont
  {Gilaberte}}, \bibinfo {author} {\bibfnamefont {M.}~\bibnamefont {Jofre}},
  \bibinfo {author} {\bibfnamefont {T.}~\bibnamefont {Scheidl}}, \bibinfo
  {author} {\bibfnamefont {J.~P.}\ \bibnamefont {Torres}}, \bibinfo {author}
  {\bibfnamefont {V.}~\bibnamefont {Pruneri}}, \ and\ \bibinfo {author}
  {\bibfnamefont {R.}~\bibnamefont {Ursin}},\ }\href {\doibase
  10.1364/JOSAB.31.002068} {\bibfield  {journal} {\bibinfo  {journal} {J. Opt.
  Soc. Am. B}\ }\textbf {\bibinfo {volume} {31}},\ \bibinfo {pages} {2068}
  (\bibinfo {year} {2014})}\BibitemShut {NoStop}%
\bibitem [{\citenamefont {Chen}\ \emph {et~al.}(2009)\citenamefont {Chen},
  \citenamefont {Pearlman}, \citenamefont {Ling}, \citenamefont {Fan},\ and\
  \citenamefont {Migdall}}]{Chen:09}%
  \BibitemOpen
  \bibfield  {author} {\bibinfo {author} {\bibfnamefont {J.}~\bibnamefont
  {Chen}}, \bibinfo {author} {\bibfnamefont {A.~J.}\ \bibnamefont {Pearlman}},
  \bibinfo {author} {\bibfnamefont {A.}~\bibnamefont {Ling}}, \bibinfo {author}
  {\bibfnamefont {J.}~\bibnamefont {Fan}}, \ and\ \bibinfo {author}
  {\bibfnamefont {A.}~\bibnamefont {Migdall}},\ }\href {\doibase
  10.1364/OE.17.006727} {\bibfield  {journal} {\bibinfo  {journal} {Opt.
  Express}\ }\textbf {\bibinfo {volume} {17}},\ \bibinfo {pages} {6727}
  (\bibinfo {year} {2009})}\BibitemShut {NoStop}%
\bibitem [{\citenamefont {Lee}\ \emph {et~al.}(2012)\citenamefont {Lee},
  \citenamefont {Kim}, \citenamefont {Cha},\ and\ \citenamefont
  {Moon}}]{Lee:2012aa}%
  \BibitemOpen
  \bibfield  {author} {\bibinfo {author} {\bibfnamefont {H.~J.}\ \bibnamefont
  {Lee}}, \bibinfo {author} {\bibfnamefont {H.}~\bibnamefont {Kim}}, \bibinfo
  {author} {\bibfnamefont {M.}~\bibnamefont {Cha}}, \ and\ \bibinfo {author}
  {\bibfnamefont {H.~S.}\ \bibnamefont {Moon}},\ }\href {\doibase
  10.1007/s00340-012-5088-4} {\bibfield  {journal} {\bibinfo  {journal}
  {Applied Physics B}\ }\textbf {\bibinfo {volume} {108}},\ \bibinfo {pages}
  {585} (\bibinfo {year} {2012})}\BibitemShut {NoStop}%
\bibitem [{\citenamefont {Jin}\ \emph {et~al.}(2015)\citenamefont {Jin},
  \citenamefont {Takeoka}, \citenamefont {Takagi}, \citenamefont {Shimizu},\
  and\ \citenamefont {Sasaki}}]{Jin}%
  \BibitemOpen
  \bibfield  {author} {\bibinfo {author} {\bibfnamefont {R.~B.}\ \bibnamefont
  {Jin}}, \bibinfo {author} {\bibfnamefont {M.}~\bibnamefont {Takeoka}},
  \bibinfo {author} {\bibfnamefont {U.}~\bibnamefont {Takagi}}, \bibinfo
  {author} {\bibfnamefont {R.}~\bibnamefont {Shimizu}}, \ and\ \bibinfo
  {author} {\bibfnamefont {M.}~\bibnamefont {Sasaki}},\ }\href {\doibase
  10.1038/srep09333} {\bibfield  {journal} {\bibinfo  {journal} {Sci Rep}\
  }\textbf {\bibinfo {volume} {5}},\ \bibinfo {pages} {9333} (\bibinfo {year}
  {2015})}\BibitemShut {NoStop}%
\bibitem [{\citenamefont {Briggs}\ \emph {et~al.}(2021)\citenamefont {Briggs},
  \citenamefont {Hou}, \citenamefont {Cui},\ and\ \citenamefont
  {Fan}}]{Briggs:21}%
  \BibitemOpen
  \bibfield  {author} {\bibinfo {author} {\bibfnamefont {I.}~\bibnamefont
  {Briggs}}, \bibinfo {author} {\bibfnamefont {S.}~\bibnamefont {Hou}},
  \bibinfo {author} {\bibfnamefont {C.}~\bibnamefont {Cui}}, \ and\ \bibinfo
  {author} {\bibfnamefont {L.}~\bibnamefont {Fan}},\ }\href {\doibase
  10.1364/OE.430438} {\bibfield  {journal} {\bibinfo  {journal} {Opt. Express}\
  }\textbf {\bibinfo {volume} {29}},\ \bibinfo {pages} {26183} (\bibinfo {year}
  {2021})}\BibitemShut {NoStop}%
\bibitem [{\citenamefont {Gayer}\ \emph {et~al.}(2008)\citenamefont {Gayer},
  \citenamefont {Sacks}, \citenamefont {Galun},\ and\ \citenamefont
  {Arie}}]{Gayer:2008aa}%
  \BibitemOpen
  \bibfield  {author} {\bibinfo {author} {\bibfnamefont {O.}~\bibnamefont
  {Gayer}}, \bibinfo {author} {\bibfnamefont {Z.}~\bibnamefont {Sacks}},
  \bibinfo {author} {\bibfnamefont {E.}~\bibnamefont {Galun}}, \ and\ \bibinfo
  {author} {\bibfnamefont {A.}~\bibnamefont {Arie}},\ }\href {\doibase
  10.1007/s00340-008-2998-2} {\bibfield  {journal} {\bibinfo  {journal}
  {Applied Physics B}\ }\textbf {\bibinfo {volume} {91}},\ \bibinfo {pages}
  {343} (\bibinfo {year} {2008})}\BibitemShut {NoStop}%
\bibitem [{\citenamefont {Dolev}\ \emph {et~al.}(2009)\citenamefont {Dolev},
  \citenamefont {Ganany-Padowicz}, \citenamefont {Gayer}, \citenamefont {Arie},
  \citenamefont {Mangin},\ and\ \citenamefont {Gadret}}]{Dolev:2009aa}%
  \BibitemOpen
  \bibfield  {author} {\bibinfo {author} {\bibfnamefont {I.}~\bibnamefont
  {Dolev}}, \bibinfo {author} {\bibfnamefont {A.}~\bibnamefont
  {Ganany-Padowicz}}, \bibinfo {author} {\bibfnamefont {O.}~\bibnamefont
  {Gayer}}, \bibinfo {author} {\bibfnamefont {A.}~\bibnamefont {Arie}},
  \bibinfo {author} {\bibfnamefont {J.}~\bibnamefont {Mangin}}, \ and\ \bibinfo
  {author} {\bibfnamefont {G.}~\bibnamefont {Gadret}},\ }\href {\doibase
  10.1007/s00340-009-3502-3} {\bibfield  {journal} {\bibinfo  {journal}
  {Applied Physics B}\ }\textbf {\bibinfo {volume} {96}},\ \bibinfo {pages}
  {423} (\bibinfo {year} {2009})}\BibitemShut {NoStop}%
\bibitem [{\citenamefont {Li}\ \emph {et~al.}(2022)\citenamefont {Li},
  \citenamefont {Mao}, \citenamefont {Weilenmann}, \citenamefont {Tavakoli},
  \citenamefont {Chen}, \citenamefont {Feng}, \citenamefont {Yang},
  \citenamefont {Renou}, \citenamefont {Trillo}, \citenamefont {Le},
  \citenamefont {Gisin}, \citenamefont {Ac\'{\i}n}, \citenamefont
  {Navascu\'es}, \citenamefont {Wang},\ and\ \citenamefont
  {Fan}}]{PhysRevLett.128.040402}%
  \BibitemOpen
  \bibfield  {author} {\bibinfo {author} {\bibfnamefont {Z.-D.}\ \bibnamefont
  {Li}}, \bibinfo {author} {\bibfnamefont {Y.-L.}\ \bibnamefont {Mao}},
  \bibinfo {author} {\bibfnamefont {M.}~\bibnamefont {Weilenmann}}, \bibinfo
  {author} {\bibfnamefont {A.}~\bibnamefont {Tavakoli}}, \bibinfo {author}
  {\bibfnamefont {H.}~\bibnamefont {Chen}}, \bibinfo {author} {\bibfnamefont
  {L.}~\bibnamefont {Feng}}, \bibinfo {author} {\bibfnamefont {S.-J.}\
  \bibnamefont {Yang}}, \bibinfo {author} {\bibfnamefont {M.-O.}\ \bibnamefont
  {Renou}}, \bibinfo {author} {\bibfnamefont {D.}~\bibnamefont {Trillo}},
  \bibinfo {author} {\bibfnamefont {T.~P.}\ \bibnamefont {Le}}, \bibinfo
  {author} {\bibfnamefont {N.}~\bibnamefont {Gisin}}, \bibinfo {author}
  {\bibfnamefont {A.}~\bibnamefont {Ac\'{\i}n}}, \bibinfo {author}
  {\bibfnamefont {M.}~\bibnamefont {Navascu\'es}}, \bibinfo {author}
  {\bibfnamefont {Z.}~\bibnamefont {Wang}}, \ and\ \bibinfo {author}
  {\bibfnamefont {J.}~\bibnamefont {Fan}},\ }\href {\doibase
  10.1103/PhysRevLett.128.040402} {\bibfield  {journal} {\bibinfo  {journal}
  {Phys. Rev. Lett.}\ }\textbf {\bibinfo {volume} {128}},\ \bibinfo {pages}
  {040402} (\bibinfo {year} {2022})}\BibitemShut {NoStop}%
\bibitem [{\citenamefont {Ho}\ \emph {et~al.}(2022)\citenamefont {Ho},
  \citenamefont {Moreno}, \citenamefont {Brito}, \citenamefont {Graffitti},
  \citenamefont {Morrison}, \citenamefont {Nery}, \citenamefont {Pickston},
  \citenamefont {Proietti}, \citenamefont {Rabelo}, \citenamefont {Fedrizzi},\
  and\ \citenamefont {Chaves}}]{Ho:2022aa}%
  \BibitemOpen
  \bibfield  {author} {\bibinfo {author} {\bibfnamefont {J.}~\bibnamefont
  {Ho}}, \bibinfo {author} {\bibfnamefont {G.}~\bibnamefont {Moreno}}, \bibinfo
  {author} {\bibfnamefont {S.}~\bibnamefont {Brito}}, \bibinfo {author}
  {\bibfnamefont {F.}~\bibnamefont {Graffitti}}, \bibinfo {author}
  {\bibfnamefont {C.~L.}\ \bibnamefont {Morrison}}, \bibinfo {author}
  {\bibfnamefont {R.}~\bibnamefont {Nery}}, \bibinfo {author} {\bibfnamefont
  {A.}~\bibnamefont {Pickston}}, \bibinfo {author} {\bibfnamefont
  {M.}~\bibnamefont {Proietti}}, \bibinfo {author} {\bibfnamefont
  {R.}~\bibnamefont {Rabelo}}, \bibinfo {author} {\bibfnamefont
  {A.}~\bibnamefont {Fedrizzi}}, \ and\ \bibinfo {author} {\bibfnamefont
  {R.}~\bibnamefont {Chaves}},\ }\href {\doibase 10.1038/s41534-022-00520-8}
  {\bibfield  {journal} {\bibinfo  {journal} {npj Quantum Information}\
  }\textbf {\bibinfo {volume} {8}},\ \bibinfo {pages} {13} (\bibinfo {year}
  {2022})}\BibitemShut {NoStop}%
\end{thebibliography}%

\end{document}